\documentclass{article}
\usepackage{spconf,amsmath,graphicx}
\usepackage{multirow}
\usepackage{lipsum}

\newcommand\blfootnote[1]{%
	\begingroup
	\renewcommand\thefootnote{}\footnote{#1}%
	\addtocounter{footnote}{-1}%
	\endgroup
}

\title{The Accented English Speech Recognition Challenge 2020: Open Datasets, Tracks, Baselines, Results and Methods}
\name{Xian Shi$^{1*}$, Fan Yu$^{1*}$\thanks{$^*$ The first two authors contributed equally to this work.}, Yizhou Lu$^2$, Yuhao Liang$^1$, Qiangze Feng$^3$, Daliang Wang$^3$, Yanmin Qian$^2$, Lei Xie$^{1\dag} $\thanks{$^\dag$ Lei Xie is the corresponding author.}}
\address{$^1$Audio, Speech and Language Processing Group (ASLP@NPU), School of Computer Science,\\ Northwestern Polytechnical University, Xi’an, China\\
	$^2$SpeechLab, Department of Computer Science and Engineering, Shanghai Jiao Tong University, China\\
	$^3$Datatang (Beijing) Technology Co., LTD, Beijing, China}
\begin{document}
	\ninept
	\maketitle
	\begin{abstract}
		
		The variety of accents has posed a big challenge to speech recognition. The Accented English Speech Recognition Challenge (AESRC2020) is designed for providing a common testbed and promoting accent-related research. Two tracks are set in the challenge -- English accent recognition (track 1) and accented English speech recognition (track 2). A set of 160 hours of accented English speech collected from 8 countries is released with labels as the training set. Another 20 hours of speech without labels is later released as the test set, including two unseen accents from another two countries used to test the model generalization ability in track 2. We also provide baseline systems for the participants. This paper first reviews the released dataset, track setups, baselines and then summarizes the challenge results and major techniques used in the submissions.
	\end{abstract}
	\vspace{-10pt}
	\section{Introduction}
	\label{sec:intro}
	\vspace{-5pt}
	
	Accent is one of the major variable factors in human speech, which poses a great challenge to the robustness of automatic speech recognition (ASR) systems. English is one of the most common languages speaking around the world. It is inevitable to produce varieties of English accents in different areas. The difference between accents is mainly reflected in three aspects of pronunciation: stress, tone and duration, which brings difficulties to ASR models.
	There has been much interest in accent recognition to distinguish different English accents~\cite{behravan2015vector,2012Speaker,najafian2020automatic,biadsy2011automatic}, and it is also valuable to improve the generalization capability of ASR models on varieties of English accents. 

	The Interspeech2020 Accented English Speech Recognition Challenge (AESRC)\footnotemark[1]\blfootnote{$^1$ \emph{~https://www.datatang.ai/interspeech2020}} is specifically designed to provide a common testbed and a sizable dataset for both English accent recognition (set as track 1) and accented English speech recognition (set as track 2). A 180-hour speech dataset is opened to participants, which contains 10 types of English accents -- Chinese, American, British, Korean, Japanese, Russian, Indian, Portuguese, Spanish and Canadian. The two tracks run on the dataset to compare the submissions fairly. 
	
	The rest of this paper is organized as follows. Section 2 is a summary of related works on accent recognition, robust speech recognition on accented speech and current datasets available for related research. Section 3 describes the dataset released by the challenge. In Section 4, the baseline experiments are introduced. Section 5 mainly summarizes the outcome of the challenge, specifically discussing on the major techniques used in the submitted systems. Section 6 concludes the challenge with important take-home messages.
	
	\vspace{-10pt}
	\section{Related work}
	\vspace{-5pt}
	Accent recognition is similar to language identification~\cite{ rangan2020exploiting, 2016Spoken, 2017Automatic} and speaker identification~\cite{okabe2018attentive,shon2018frame,xie2019utterance,nagrani2020voxceleb}. They all classify variable-length speech sequences to utterance-level posteriors to obtain accent, speaker or language ID. In order to distinguish different accents in English, Teixeira et al.~\cite{teixeira1996accent} proposed to use context-dependent HMM units to optimize parallel networks and Deshpande et al.~\cite{deshpande2005accent} introduced format frequency features into GMM models. 
	Ahmed et al.~\cite{ahmed2019vfnet} presented VFNet (Variable Filter Net), a convolutional neural network (CNN) based architecture which applies filters with variable size along the frequency band to capture a hierarchy of features, aiming at improving the accuracy of accent recognition in dialogues. Winata et al.~\cite{winata2020learning} proposed an accent-agnostic approach that extends the model-agnostic meta-learning (MAML) algorithm for fast adaptation to unseen accents. Transfer learning and multitask learning were also found useful for spoken accent recognition tasks~\cite{meng2020vector,viglino2019end}.
	
	\begin{table*}[htbp] 
		\caption{Results of baseline systems on the separated cv set.}
		\centering 
		\vspace{-0.1cm}
		\includegraphics[width=0.95\linewidth,scale=1.0]{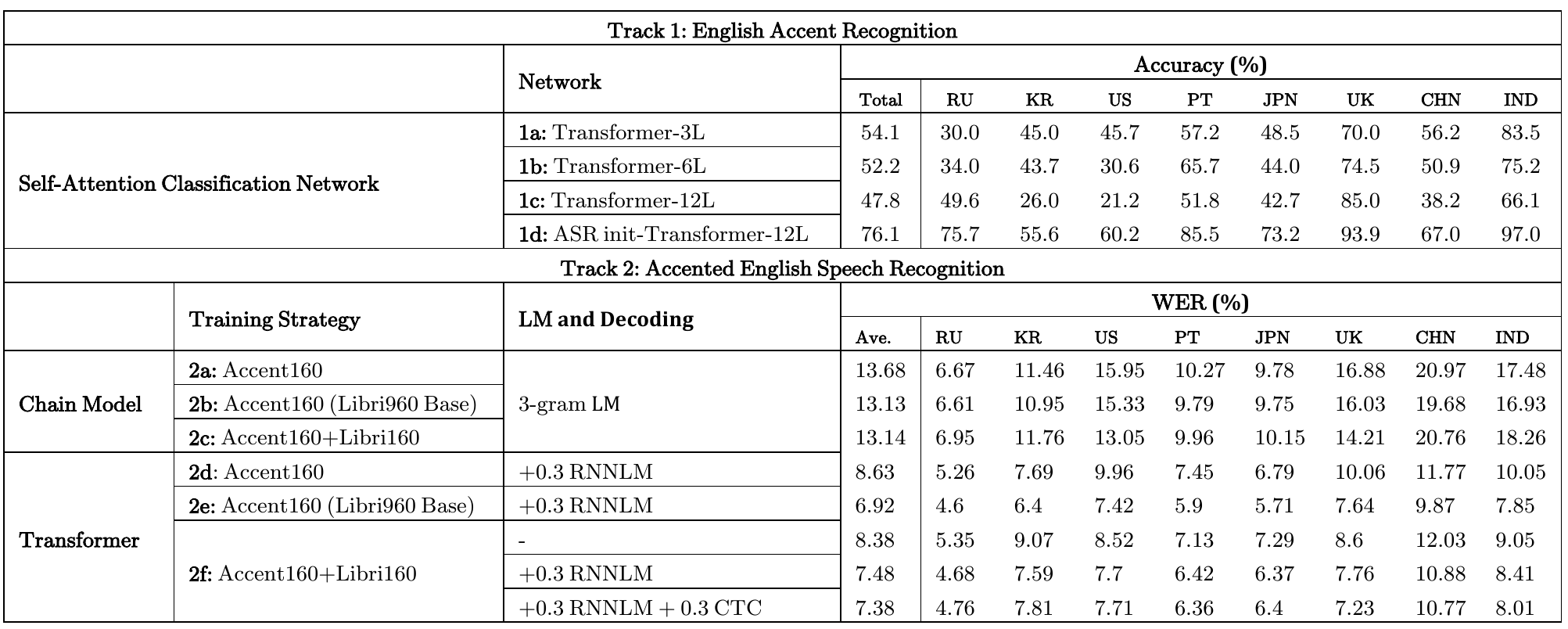} 
		\label{base}
		\vspace{-0.4cm}
	\end{table*}
	
	As for speech recognition on accented speech, adaptation methods and adversarial training related techniques were proved effective. Assuming that the labelled data for specific accent is limited, the adaptation method is first to train a base model with standard speech data that is usually with a large volume, and then adapt the model to the specific accent  with the respective data~\cite{li2010comparison,zhu2020multi,liao2013speaker,tan2015cluster}. Domain adversarial training (DAT) was used by Sun et al. to obtain accent-independent hidden representation in order to achieve a high-performance ASR system for accented Chinese~\cite{sining}. A generative adversarial network (GAN) based pre-training framework named as AIPNet was proposed by Chen et al.\cite{chen2019aipnet}. They pre-trained an attention-based encoder-decoder model to disentangle accent-invariant and accent-specific characteristics from acoustic features by adversarial training. 
	Accent-dependent acoustic modeling approaches take accent-related information into network architecture by accent embedding, accent-specific bottleneck features or i-vectors~\cite{2019A, Chen2015ImprovingDN}. 
	In a closed set of known accents, accent-dependent models usually outperform the accent-independent universal models, while the latter ones usually achieve a better average model under the situations where accent labels are unavailable.
	
	English accent recognition and accented English speech recognition are also hindered by data insufficiency. Existing open-source accented English datasets are limited in data volume and accent varieties~\cite{demirsahin-etal-2020-open,magic-data-2019}. This motivates us to provide a sizable dataset and a comon testbed to advance the research in the related areas.


	\section{Open Dataset}
	
	An accented English speech dataset was released to participants in the challenge. It was collected from both native speakers in UK and US, and also English speakers in China, Japan, Russia, India, Portugal, Korea, Spain and Canada. We suppose that the data collected in each country is belong to one type of accent of English, and in total we have 10 `accents'. The speakers, aged between 20 to 60, were asked to read sentences covering common conversation and human-computer speech interaction commands. All the speech recordings were collected in relatively quiet indoor environment with Android phones or iPhones. The training set, named as Accent160, for both challenge tracks (introduced in Section 4) has 160 hours of speech including 8 accents (20 hours/accent). Spanish and Canadian accents are not included in the training set. The test set for track 1 includes 16 hours of data (2 hours for each accent) and the test set for track 2 has 20 hours of data including extra 4 hours of data from Spanish and Canadian accents (as unseen accent data). Speech recordings were provided in Microsoft wav format (16KHz, 16bit and mono) with manual transcriptions.
	
	Training speech data and the corresponding transcriptions were first released to participants together with metadata, in which detailed information about speakers and recording environment including speaker gender, age, region, recording device and others are provided. In order to make a fair comparison with the provided baseline experiments, we also release a speaker list for participants to divide the CV set from the training set. The test set was released later to the participants with only audio recordings.
	
	\begin{table*}[htbp] 
		\caption{Results and major techniques used in the top 8 submissions in track 1.}
		\centering 
		\vspace{0.1cm}
		\includegraphics[width=0.95\linewidth,scale=1.00]{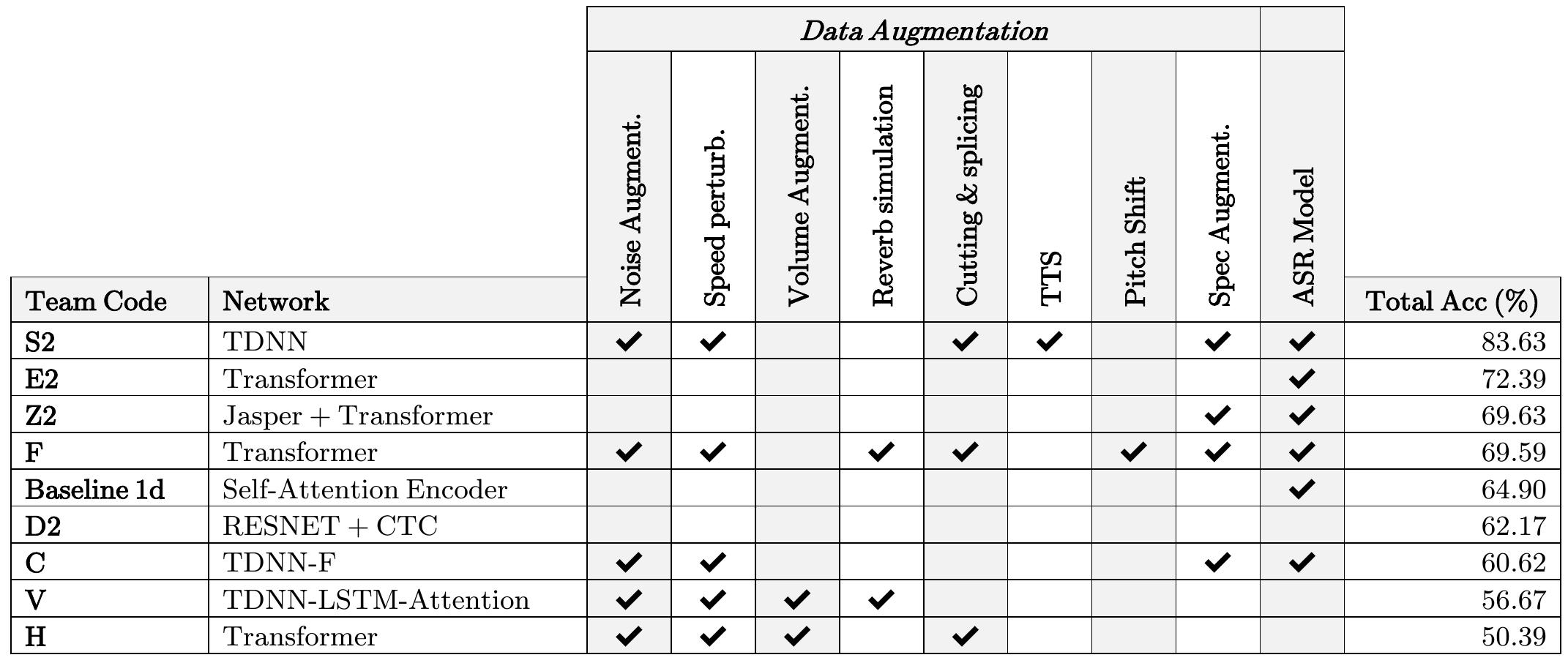} 
		\label{track1}
		\vspace{-0.5cm}
	\end{table*}
	\vspace{-0.2cm}
	\section{Tracks and Baselines}
	
	\subsection{Track 1: English Accent Recognition}
	Track 1 aims to study the English accent recognition problem. The rules are as follows. 1) Data used for accent classification training is limited to the 160 hours Accent English data and 960 hours of Librispeech~\cite{panayotov2015librispeech} data, and data augmentation based on the above data is permitted. 2) Multi-system fusion techniques including recognizer output voting error reduction (ROVER)~\cite{fiscus1997post} are prohibited; there is no other restriction to the model and training techniques. 3) The final ranking is based on the recognition accuracy on the whole test set and the accuracy for each accent is for reference only.
	
	For baseline experiments, a self-attention (SA) based classification network is realized using ESPnet\footnote[2]{\emph{~https://github.com/espnet/espnet}}. A mean + std pooling layer is applied after encoder to pool the output on \emph{T} dimension. 
	\emph{Transformer-3L$\backslash$6L$\backslash$12L} are different in the number of encoder layers. All of them are trained under the simple CE loss for 40 epochs. Specaugment is applied to the input feature. As shown in Table~\ref{base}, with the use of the releasd data, 6$\backslash$12 layers of SA encoder result in over fitting. From the results on the CV set, we found that the accuracy of some accents varies a lot among different speakers. As there are only a few speakers in the CV set, the absolute value above is not statistically significant. However, it is worth noting that when we use an SA encoder trained by an ASR downstream task to initialize the encoder of the accent classification network, the accent recognition accuracy is significantly improved. Finally the total accuracy of ASR-init-Transformer-12L is up to 76.1\% on the CV set. The code and configuration of the baseline can be found from our github\footnote[3]{\emph{~https://github.com/R1ckShi/AESRC2020}}.

	\subsection{Track 2: Accented English ASR}
	
	Track 2 studies the robustness of ASR system on accented English speech where the word error rate on the whole test set is used as the evaluation metric. Test sets include accents beyond training data in order to evaluate the generalization performance of the model. Again, data usage is only limited to the released data and the librispeech data. All kinds of system combination methods including ROVER are strictly prohibited. Language model training should only use the transcripts of permitted speech training data. Data augmentation should only be applied on the permitted speech data only.
	
	We prepare ASR baseline systems for track 2 with both Kaldi\footnote[4]{\emph{~http://www.kaldi-asr.org/}} and ESPnet toolkits. Several training strategies and decoding related parameters are compared, and results are shown in the Table~\ref{base}.  
	
	In all experiments, we use 71-dimensional mel-filterbank feature as the input of the acoustic model and frame length is 25 ms with a 10 ms shift. In our baseline chain-model system, the acoustic model consists of a single convolutional layer with the kernel of 3 to down-sample the input speech feature, 19 hidden layers of a 256-dimensional TDNN and a 1280-dimensional fully connection layer with ReLU activation function. A 3-gram language model trained with the transcripts of the 160 hours of speech is used in the decoding graph compiling.
	As for the transformer baseline, ESPnet joint CTC/Attention transformer which contains 12-layer encoder and 6-layer decoder is used, and the dimension of attention and feed-forward layer is set to 256 (4 heads) and 2048 respectively. The whole network is trained for 50 epochs with warmup\cite{DBLP:journals/corr/VaswaniSPUJGKP17} for the first 25,000 iterations. We mainly try three training strategies: 1) only using Accent160; 2) using Accent160 with another 160 hours of selected data from librispeech (Libri160); 3) using 960 hours of librispeech data (Libri960) to train a base model and then fine-tuning the model using Accent160. Furthermore, we optimize the decoding with RNN language model and CTC posterior probability. The RNNLM is a 2-layer LSTM model trained using ESPnet on the transcriptions associated with Accent160, and both RNNLM and CTC are fused (weight=0.3 for both) with beam search scores. From the baseline results, we find that the end-to-end models outperform the hybrid chain models given the limited training data and fine-tuning the librispeech base model with accented English data achieves the best performance among the three training strategies. The whole recipe and  results of the baselines can be found from our github.	
	\begin{table*}[htbp] 
		\caption{Results and major techniques used in the top 10 submissions in track 2.}
		\centering 
		\vspace{0.1cm}
		\includegraphics[width=0.95\linewidth,scale=1.00]{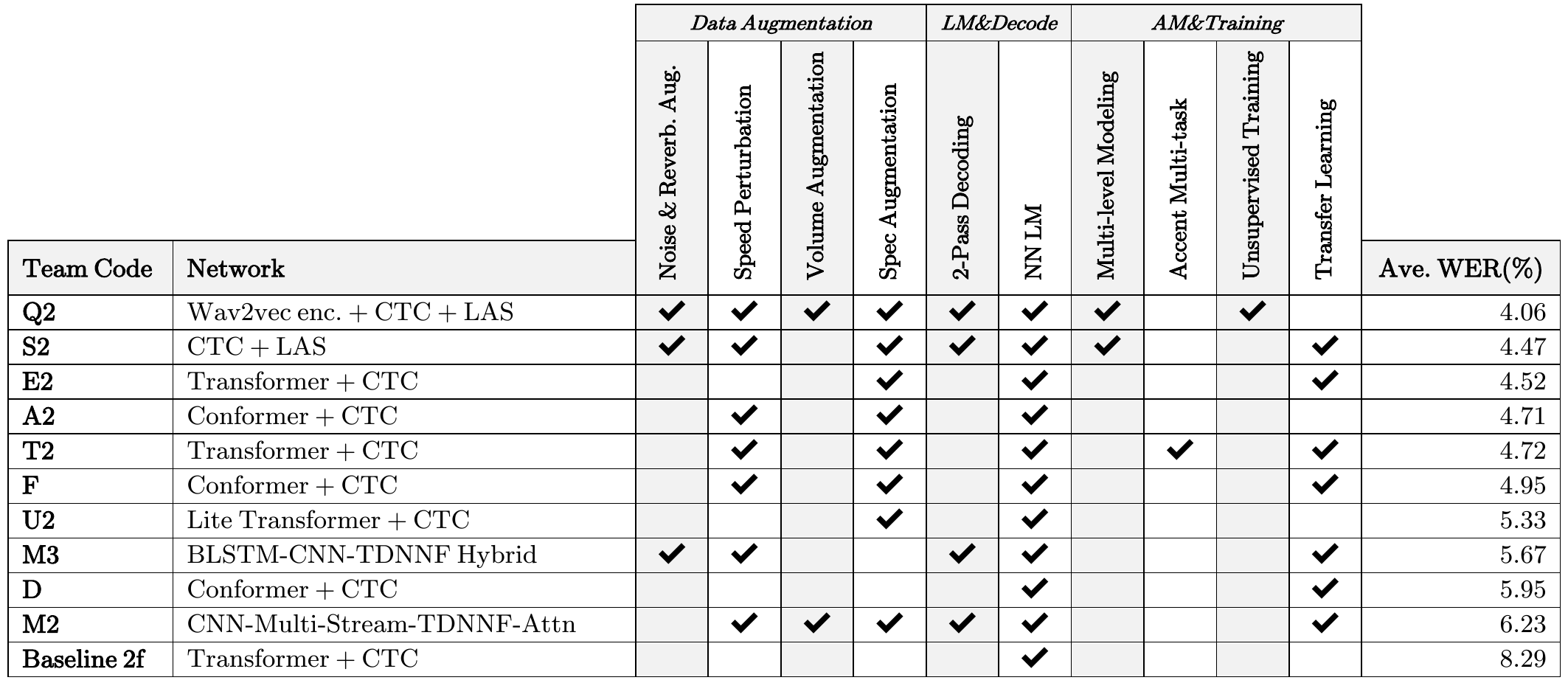}
		\label{track1}
		\vspace{-0.5cm}
	\end{table*}
	\vspace{-0.cm}
	\section{Challenge Results and Analysis}
	\label{sec:pagestyle}
	\subsection{Track 1: English Accent Recognition}
	Finally 25 teams submitted their results to track 1 and the accuracy on the test set and the major techniques used for the top 8 teams are summarized in Table~\ref{track1}. The winner goes to team S2 using a TDNN based classification network with phonetic posteriorgram (PPG) feature as input, and they use text-to-speech (TTS) to expand the training data~\cite{sjtu-accent}. The major techniques are summarized below.
	\vspace{-0.25cm}
	\subsubsection{Data augmentation}
	\vspace{-0.05cm}
	Since the size of the released training data is relatively small, most teams have done a lot of work in data augmentation. For example, noise augmentation and speed perturbation are generally used. 
	Speed augmentation can enhance the robustness of the model modestly, while volume augmentation and reverberation simulation help a little. Simulated room impulse response (RIR) is used to convolution with the original speech to generate data with reverberation. Moreover, several teams use random cutting and splicing to expand the data. In detail, two pieces of audio with the same accent are randomly selected from the training dataset, and then each piece is cut to two splices and the splices from the two pieces are combined as new samples. Specaugmentation is also very useful reported by many teams. Pitch shift is also an effective data augmentation method reported by team F. In addition to the above tricks, it is also worth to notice that the winner team S2 used the provided data to train a TTS system to synthesize a large number of training audio clips with the corresponding accents, and the accuracy of accent recognition was improved absolutely by 10\%.
	
	\vspace{-0.25cm}
	\subsubsection{Training Strategy}
	\vspace{-0.05cm}
	As revealed in the baseline experiments, the training of accent classification is easy to be over fitted to the speakers as the biggest acoustic difference lies in the speaker characteristics rather than accent characteristics, which leads to great difference in the accuracy of different speakers with same accent. Therefore, it is beneficial to use speaker-invariant feature input and pre-trained encoder by speaker-invariant downstream task to initialize the network. Team S2 used PPG features generated by a Kaldi ASR system as model input. Team Z2 adopted a multi-task strategy with both accent recognition and phoneme classification. The second place team E2 used the accent label together with transcripts to train a Transformer ASR model with accent classification ability at the same time. As reported, putting the accent tag at the beginning of text outperforms tagging at the end.
	Besides the mainstream neural networks, team H used an NN and support vector machine (SVM) combination method. An embedding layer was applied before the fully-connected layer and SVM was used to classify the embedding vector.
	
	\vspace{-0.2cm}
	\label{sec:typestyle}
	\subsection{Track 2: Accented English ASR}
	
	Forty two teams submitted their results to track 2. Team Q2 obtained the lowest average WER of 4.06\%. This was achieved by a CTC model with LAS rescoring while the CTC model was initialized by a Wav2Vec encoder trained in an unsupervised manner using Fairseq toolkit~\cite{ott2019fairseq}. The superior performance indicates that unsupervised training is promising in improving performance when labeled data is limited. The results and major techniques used in the top 10 systems are summarzied in Table 3.
	
	\vspace{-0.2cm}
	
	\subsubsection{Data augmentation}
	
	Similar to track 1, various data augmentation tricks were widely adopted in the submitted system in track 2 and Table 3 shows the tricks used in the top-performing systems. According to the system descriptions provided by the teams, the relative WER reduction of 5\% to 10\% can be achieved by methods including volume augmentation and speed perturbation. Noise and reverberation augmentation was tried by several teams but no obvious gain was obtained. This is probably because the acoustic and channel conditions between the test set and the training set is similar.
	
	\vspace{-0.2cm}
	
	\subsubsection{Network Structure}
	A variety of different models were found to be used in track 2, mainly including Transformer-based encoder-decoder models~\cite{vaswani2017attention, 2020Conformer, 2020Lite}, CTC models with LAS~\cite{chan2016listen} rescoreing and traditional hybrid models. Attention-based end-to-end models are able to take full sentence-scale acoustic information into consideration, so they have a natural advantage over the traditional hybrid models. Unsupervised training has been drawing increasingly attention~\cite{2020wav2vec, xingchen2019speech-xlnet}. Team Q2 followed the work of wav2vec2.0~\cite{2020wav2vec} and pre-trained a self-attention encoder using both contrastive loss and diversity loss, in order to obtain an encoder with waveform reconstruction capability and contextualized representation capability. The score of letter-level model above was combined with a word-piece level Transformer LM. The second place team S2 adopted frame-level CE loss pre-trained encoder~\cite{tan2021aispeech} (labels are generated by a Kaldi triphone system), resulting in 5\% WER reduction. Conformer and lite Transformer were used by several teams which implies the potential room for improvement of primitive Transformer, especially in enhancing the ability of local information modeling. Explicit accent related optimizations were rarely used in the submitted systems. But team T2 used accent recognition multi-task training encoder and yielded 3\% relative WER reduction.
	
	\vspace{-0.2cm}
	\subsubsection{Language Modeling}
	
	As for language modeling, it is obvious that NN language model rescoring works well, and it brings improvement ranging from 7\% to 15\% on the CV set, reported from the system descriptions. Two-pass decoding was used by the top 2 teams. It is effective to rescore the lattice generated by WFST (11.7\% WER reduction by team S2) or fusion the primitive posterior probability in the process of beam search with LAS (26.4\% WER reduction by Q2). 
	Team Q2 specifically compared the performance of statistic language model and NN language model. It turns out that a well-trained Transformer LM can achieve a slightly lower perplexity, but the 4-gram model outperforms the Transformer LM in WER (3.96\% to 4.01\%). Team Q2 also tried a method of combining two language models using different granularity modeling units. A word-piece level Transformer LM was applied in the decoding fusion, leading to 2.4\% relative WER reduction on the CV set (3.73\% to 3.64\%).
	
	\vspace{-0.1cm}
	\section{Summary}
	In this challenge, participants have used the released 160 hours of training data to build accent recognition (track 1) and accented English speech recognition (track 2) systems. According to the results of track 1, we found it necessary to fix the over-fitting problem, which means to peel off the speaker-related information from the encoder output. Therefore, a pre-trained encoder with ASR downstream task works well. The use of phonetic posteriorgram (PPG) features as network input is also effective in accent recognition. 	
	In track 2, we have seen a variety of networks including end-to-end models and traditional hybrid systems. In conclusion, CNN-based unsupervised training with contrastive loss and diversity loss can enhance the ability of waveform reconstruction and contextualized representation of encoder, leading to superior recognition performance. 
	Several works are done on the Transformer family like Conformer and Lite Transformer. The oracle Transformer has disadvantages in local information modeling and strengthening such local information apparently leads to improved performance. As to language modeling, CTC with LAS 2-pass decoding performs well, which combines the time sequence modeling capability with CTC loss and full sentence scale modeling ability from self-attention structures. A few accent related modeling techniques are used in track 2, but most participants choose to train an average model for all accents. With limited training data, data augmentation tricks are essential for both tracks.

	
	
	\bibliographystyle{IEEEbib}
	\bibliography{strings, refs}
	
\end{document}